\documentclass[%
 aip,
 rsi,
 amsmath,amssymb,
 reprint,%
]{revtex4-1}

\usepackage{epsfig}
\usepackage{natbib}
\usepackage{color}
\usepackage{amsmath}
\usepackage{amssymb}
\usepackage{verbatim} 

\usepackage{graphicx}
\usepackage{dcolumn}
\usepackage{bm}

\newcommand{\beq}{\begin{equation}}
\newcommand{\eeq}{\end{equation}}
\newcommand{\vv}{\mathbf{v}}

\newcommand{\bvec}{\begin{pmatrix}}
\newcommand{\evec}{\end{pmatrix}}
\newcommand{\lp}{\left(}
\newcommand{\rp}{\right)}

\newcommand{\eps}{\epsilon}

\newcommand{\pa}[2]{\frac{\partial #1}{\partial #2}}
\newcommand{\tot}[2]{\frac{d #1}{d #2}}

\newcommand{\Bth}{B_\theta}
\newcommand{\bth}{b_\theta}
\newcommand{\pp}{p_{\perp i}}
\renewcommand{\pl}{p_{\parallel i}}

\newcommand{\Pv}{\mathbf{P}}
\newcommand{\Rv}{\mathbf{R}}
\newcommand{\Ev}{\mathbf{E}}
\newcommand{\Bv}{\mathbf{B}}
\newcommand{\ve}[1]{\mathbf{#1}}
\newcommand{\pps}[1]{p_{\perp {#1}}}
\newcommand{\pls}[1]{p_{\parallel {#1}}}
\newcommand{\tn}{\tilde{ n}}
\newcommand{\tr}{\tilde{ r}}
\renewcommand{\tt}{\tilde{ t}}

\begin{document}
	

\title[Anisotropy-driven collisional separation of impurities in magnetized compressing and expanding cylindrical plasmas]{Anisotropy-driven collisional separation of impurities in magnetized compressing and expanding cylindrical plasmas}

\author{I. E. Ochs}
\affiliation{Department of Astrophysical Sciences, Princeton University, Princeton, New Jersey 08540}
\affiliation{Princeton Plasma Physics Laboratory, Princeton, New Jersey 08543}  

\author{N. J. Fisch}
\affiliation{Department of Astrophysical Sciences, Princeton University, Princeton, New Jersey 08540}
\affiliation{Princeton Plasma Physics Laboratory, Princeton, New Jersey 08543}

\date{\today}

\begin{abstract}
	When a cylindrically-symmetric magnetized plasma compresses or expands, velocity-space anisotropy is naturally generated as a result of the different adiabatic conservation laws parallel and perpendicular to the magnetic field.
	When the compression timescale is comparable to the collision timescale, and both are much longer than the gyroperiod, this pressure anisotropy can become significant.
	We show that this naturally-generated anisotropy can dramatically affect the transport of impurities in the compressing plasma, even in the absence of scalar temperature or density gradients, by modifying the azimuthal frictions that give rise to radial particle transport.
	Although the impurity transport direction depends only on the sign of the pressure anisotropy, the anisotropy itself depends on the pitch magnitude of the magnetic field and the sign of the radial velocity.
	Thus, pressure anisotropy effects can drive impurities either towards or away from the plasma core.
	These anisotropy-dependent terms represent a qualitatively new effect, influencing transport particularly in the sparse edge regions of dynamically-compressing screw pinch plasmas.
	Such plasmas are used for both X-ray generation and magneto-inertial fusion, applications which are sensitive to impurity concentrations.
\end{abstract}

\maketitle





\section{Introduction}

In both inertial and magnetic confinement fusion reactors, highly-charged impurities from the wall can penetrate into the core plasma, choking the fusion reaction.
The presence of a magnetic field, introduced to improve heat confinement, can vastly change the impurity transport.
In the unmagnetized case, relevant for inertial fusion, impurities tend to be pushed into hot, dense regions of the plasma \cite{vekshtein1975diffusion, amendt2010barodiffusion, kagan2014thermodiffusion}.
In the magnetized case, relevant for steady-state magnetic fusion, impurities tend to be relegated to cold, dense regions of the plasma \cite{spitzer1952equations,taylor1961diffusion,hirshman1981neoclassical}.
Recently, the development of magnetized inertial fusion concepts such as magnetized liner inertial fusion (MagLIF) \cite{slutz2010pulsed, ryutov2015characterizing} and magnetic target fusion \cite{intrator2004high,intrator2004high2} has made magnetized transport processes relevant to the inertial fusion community as well, since they can be harnessed to reduce fuel pollution due to impurities \cite{vekshtein1975diffusion, ryutov2015characterizing, garcia2018mass} and ash \cite{ochs2018favorable} in the fusion core.

Because magneto-inertial fusion involves the compression of a magnetic field, it can introduce additional complications to the transport.
In particular, a compressing magnetized plasma naturally generates pressure anisotropy, due to the conservation of adiabatic invariants parallel and perpendicular to the field\cite{kulsrud1980mhd}.
When the magnetic field is axial, the perpendicular pressure increases faster than the parallel pressure, while the converse is true in an azimuthal-magnetic-field plasma.
Since this anisotropy can alter the divergence of the pressure tensor, it can also alter the diamagnetic flows and frictions that determine the impurity transport in a magnetized plasma.

In this paper, we show that the generated anisotropy is non-negligible when the compression rate is a substantial fraction of the collisional isotropization rate, and that this anisotropy can significantly impact the transport when the magnetic field lines are curved.
For example, a test case with approximate low-density laboratory screw pinch parameters yields a transient 40\% deviation in impurity density from the isotropic-pressure case.

Although these conditions are not relevant to transport in the core region of MagLIF \cite{ochs2018favorable}, they could occur in more dilute experiments or regions of the plasma.
For instance, the axial current in MagLIF results in a curved magnetic field in the rarified edge regions of the plasma.
In addition, Z-pinch experiments have employed helical return-current wires to stabilize the implosions\cite{sorokin2013gas,zukakishvili2005}, and a similar strategy has been proposed for MagLIF\cite{schmit2016controlling}.
Such return-current wire arrays produce a dynamically-compressing screw pinch outside the core plasma, in a region which could contain many different density and temperature regimes, including those which support temperature anisotropy.

In general, solving for the multi-fluid extended MHD dynamics of compression is a complex problem.
However, we are primarily interested in the \textbf{relative} radial motion of the impurity and the background.
In a low-Mach plasma, this relative radial motion consists primarily of an $F \times B$ drift arising from the azimuthal friction force between the species.
This azimuthal friction arises from the difference in the species' diamagnetic drifts, which depends both on the density gradients and pressure anisotropy within the plasma. 
These density gradients and anisotropies should evolve in a qualitatively similar way for a wide variety of compression profiles.
Thus, rather than solve for the perpendicular collisional transport with self-consistent electric fields, we instead choose electric and magnetic fields which make our analytical calculation straightforward and tractable.
Specifically, we choose a magnetic field with a constant rotational transform and uniform initial magnitude, and an electric field consistent with self-similar exponential compression.
By making these simplifying assumptions, we elucidate the qualitatively new transport effects that result from pressure anisotropy generated in the compressing plasma.

We therefore begin in Section \ref{sec:transport} by deriving the transport equations in a slowly compressing plasma with an imposed anisotropy.
This allows us to determine the tendency of anisotropic effects to draw impurities towards the plasma center vs. towards the edge, as a result of radial $F \times B$ drifts from the anisotropy-dependent azimuthal friction forces.
Then in Section \ref{sec:anisotropy}, we derive the anisotropy that naturally arises in the compressing plasma, by combining the double-adiabatic MHD closure with an anisotropy relaxation model.
In particular, we show how compressing plasmas with strong axial magnetic fields tend to have larger perpendicular pressure, while those with strong azimuthal fields tend to have larger parallel pressure, and how the converse is true during plasma expansion.
In Section \ref{sec:quasistationary}, we combine this anisotropy model with our transport model and take the limit of infinitely fast diffusion to yield the quasi-stationary state, i.e. the impurity distribution towards which the plasma naturally tends to evolve at a given snapshot in time during the compression.
Finally, in Section \ref{sec:dynamics}, we consider the full dynamical problem, transforming the impurity diffusion equation into a compressing frame, to elucidate how the transport dynamics naturally change over the course of the compression.

\section{Ion transport model} \label{sec:transport}

We begin by deriving the transport equations in a cylindrically symmetric compressing plasma from the fluid momentum equations, allowing for pressure anisotropy.
We will then show how anisotropy is naturally generated in the plasma from the adiabatic equations.
By combining these models, we will show that the tendency of impurities to accumulate in the plasma core depends both on the direction of radial acceleration and the curvature of the magnetic field.

Consider a cylindrically-symmetric, isothermal, multi-species plasma, with $B_r = 0$.
The plasma contains a bulk species $b$, and a trace high-charge impurity $s$.
Each species $i \in \{b,s\}$ obeys the fluid momentum equation:
\beq
n_i m_i \tot{\vv_i}{t} = n_i Z_i e \lp \Ev + \vv_i \times \Bv \rp -\nabla \cdot \Pv_i + \sum_j \Rv_{ij}, \label{eq:momentum}
\eeq
where the friction force is given by
\beq
\Rv_{ij} = \frac{m_i n_i}{\tau_{ij}} (\vv_j - \vv_i) \label{eq:friction},
\eeq
and the momentum exchange collision time is given by \cite{braginskii1965transport}
\beq
	\tau_{ij} = \frac{3}{4\sqrt{2\pi}} \frac{m_i}{\sqrt{m_{ij}}} \frac{T^{3/2}}{\lambda_{ij} e^4 Z_i^2 Z_j^2}.
\eeq
Here, $m_{ij} \equiv m_i m_j / (m_i + m_j)$ is the reduced mass, $\lambda_{ij}$ is the Coulomb logarithm, and $Z_i$ is the charge state of ion $i$.

Now, define the direction $\hat{\wedge} = \hat{r} \times \hat{b}$.
We assume 
\beq
\Ev = E_r \hat{r} + E_\wedge \hat{\wedge},
\eeq
so that there are no electric fields along the field lines.
Then, crossing Eq.~(\ref{eq:momentum}) with $\hat{b}$ yields:
\begin{align}
\vv_i &= -\frac{E_\wedge}{B} \hat{r} + \frac{E_r}{B} \hat{\wedge} - \frac{1}{\Omega_i} \frac{\lp \nabla \cdot \Pv_i \rp \times \hat{b}}{n_i m_i} \notag\\
&\qquad +\frac{1}{\Omega_i \tau_{ij}} \lp \vv_j - \vv_i \rp \times \hat{b} -\frac{1}{\Omega_i} \tot{\vv_i}{t} \times \hat{b}. \label{eq:vperp}
\end{align}
This equation describes dynamics purely perpendicular to a helical magnetic field.

Now, we order this equation for magnetized transport by taking the following as small:
\begin{align}
\frac{1}{\Omega_i \tau_{ij}} &\ll 1\\
\frac{1}{\Omega_i} \frac{1}{v_i} \left| \tot{\vv_i}{t} \right| &\ll 1. \label{eq:inertiaConstraintGeneral}
\end{align}
We will make the second ordering more explicit later, when we examine the constituent terms in the total velocity derivative. 
With this ordering, Eq. (\ref{eq:vperp}) becomes:
\begin{align}
\vv_i^{(0)} &= -\frac{E_\wedge}{B} \hat{r} + \frac{E_r}{B} \hat{\wedge} - \frac{1}{\Omega_i} \frac{\lp \nabla \cdot \Pv_i \rp \times \hat{b}}{n_i m_i} \label{eq:v0}\\ 
\vv_i^{(1)} &= \frac{1}{\Omega_i \tau_{ij}} \lp \vv_j^{(0)} - \vv_i^{(0)} \rp \times \hat{b} -\frac{1}{\Omega_i} \tot{\vv_i^{(0)}}{t} \times \hat{b} \label{eq:v1}\\
&= -\frac{1}{m_i \Omega_i^2 \tau_{ij}} \lp \frac{Z_i}{Z_j} \frac{\nabla \cdot \Pv_j }{n_j}  - \frac{\nabla \cdot \Pv_i  }{n_i} \rp \times \hat{b} \times \hat{b} \notag \\
& \qquad -\frac{1}{\Omega_i} \tot{\vv_i^{(0)}}{t} \times \hat{b}. \label{eq:v1a}
\end{align}
The 0th-order motion consists of $E \times B$ and diamagnetic drifts, whereas the 1st-order motion contains the friction-driven transport and inertial effects.

The transport dynamics greatly simplify if the first order bulk ion radial velocity vanishes, i.e. $v_{br}^{(1)} = 0$.
To this end, we assume that the impurity forms a small enough fraction of the overall plasma that $\tau_{bs} \rightarrow \infty$, which allows us to neglect the first term in Eq. (\ref{eq:v1}).
To eliminate the inertial terms, we first expand the total derivative of the velocity:
\begin{align}
\frac{1}{\Omega_i} \tot{\vv_i^{(0)}}{t} \times \hat{b} &= \frac{1}{\Omega_i} \biggl[ \lp \tot{v_{ir}^{(0)}}{t} - \frac{\lp v_{i\theta}^{(0)}\rp^2}{r} \rp \hat{r} \notag \\
& \qquad  +\lp \tot{v_{i\theta}^{(0)}}{t}  + \frac{v_{ir}^{(0)} v_{i\theta}^{(0)}}{r} \rp \hat{\theta} \notag\\
& \qquad + \lp \tot{v_{iz}^{(0)}}{t} \rp \hat{z}\biggr] \times \hat{b}. \label{eq:totalDeriv}
\end{align}
Here, we have rewritten the total (Lagrangian) derivative of the velocity vector in terms of the total derivative of its components.

We now make our main simplification.
We take a given form of $\Ev$ which leads to analytically tractable results, so as to more clearly elucidate the transport mechanisms involved.
Limitations of this model will be addressed in the discussion section.

To choose $\Ev$, we first note that if $v_{i\theta}^{(0)} = 0$ and $v_{iz}^{(0)} = 0$, then the $\hat{r}$ component of Eq.~(\ref{eq:totalDeriv}) (given by the second two terms in the brackets).
Thus to eliminate inertial effects on the radial transport of the bulk, we take 
\beq
E_r \equiv \frac{B}{\Omega_b} \frac{\lp \nabla \cdot \Pv_b \rp \times \hat{b}}{n_b m_b}. \label{eq:Er}
\eeq
This ensures that the non-$\hat{r}$ components of Eq.~(\ref{eq:v0}) cancel for species $b$.
Second, because it leads to an analytically simple form of compression, we take $E_\wedge = k r B$, leaving us with the equation (valid to first order):
\beq
v_{br} =v_{br}^{(0)} + v_{br}^{(1)} =  -kr. \label{eq:vbr}
\eeq
The subsequent analysis could be extended to other self-similar compression profiles \cite{velikovich1985hydrodynamics,ramsey2010class} by simply taking a time-dependent $k(t)$, as long as $dk/dt \ll 1$.

With our bulk species dynamics established, we turn to the impurity $s$.
First, we note that $\nabla \cdot \Pv_s \parallel \hat{r}$ for any diagonal pressure tensor that is a function only of $r$.
Second, we note that $v_{s\theta}^{(0)} = -b_z v_{s\wedge}^{(0)}$, and $v_{sz}^{(0)} = \bth v_{s\wedge}^{(0)}$.
Thus, plugging our explicit electric field forms into Eqs.~(\ref{eq:v0}-\ref{eq:v1}) and noting $v_{b \wedge} = 0$, we find:
\begin{align}
\vv_s^{(0)} &= -kr \hat{r} + \lp  \frac{1}{\Omega_b} \frac{\lp \nabla \cdot \Pv_b \rp \cdot \hat{r}}{n_b m_b} - \frac{1}{\Omega_s} \frac{\lp \nabla \cdot \Pv_s \rp \cdot \hat{r}}{n_s m_s} \rp  \hat{\wedge}\\ 
\vv_s^{(1)} &= \frac{1}{\Omega_s \tau_{sb}} \lp  v_{s\wedge}^{(0)} \rp \hat{r} 
\notag\\
&\quad + \frac{1}{\Omega_s} \biggl[ - \lp \tot{v_{sr}^{(0)}}{t} - b_z^2 \frac{\lp v_{s\wedge}^{(0)}\rp^2}{r} \rp \hat{\wedge} \notag\\
&\qquad \qquad +\lp \tot{v_{s\wedge}^{(0)}}{t} - b_z^2 k  v_{s\wedge}^{(0)} \rp \hat{r}  \biggr] \label{eq:v1s} . 
\end{align}

We adopt the ordering:
\begin{align}
\left | \frac{k}{\Omega_s} \frac{k r}{v_{s\wedge}} \right| &\ll k \tau_{sb}\\
\left | b_z^2 \frac{v_{s\wedge}^{(0)}}{\Omega_s r} \right | &\ll k \tau_{sb},
\end{align}
This ordering ensures that inertial effects from rotation and compressional acceleration are negligible compared to the pressure anisotropy effects.
For details on the effect of non-negligible plasma rotation, see e.g. Kolmes \emph{et al}\cite{kolmes2018strategies}.


For the calculation of $v_{s\wedge}^{(0)}$, we use a model for the pressure tensor with different pressures parallel and perpendicular to the magnetic field, i.e.
\beq
\Pv_i = \pp \ve{I} + (\pl - \pp) \hat{b}\hat{b} .
\eeq
The divergence of this tensor is given by
\beq
\nabla \cdot \Pv_i = \pa{\pp}{r}  + \lp \pp - \pl \rp \frac{\bth^2}{r}.
\eeq

Thus, assuming $T_\perp$ is spatially constant and the same for each species, the $\hat{\wedge}$ compenent of the velocity is:
\begin{align}
v_{s\wedge}^{(0)} &= - \frac{T_\perp}{ m_s \Omega_s} \left[   \frac{\partial_r n_s}{n_s} -\frac{Z_s}{Z_b} \frac{\partial_r n_b}{n_b} + \lp \eps_s - \frac{Z_s}{Z_b} \eps_b \rp \frac{\bth^2}{r}    \right]. \label{eq:vwedge}
\end{align}
where $\eps_s \equiv (T_{\perp s}  - T_{\parallel s}) / T_{\perp s}$ is a parameter describing the temperature anisotropy.
The first two terms, in a scalar-pressure plasma, give rise to the well-known impurity pinch effect\cite{taylor1961diffusion}.
The last term is new, and describes the effect of pressure anisotropy on the transport dynamics.

From the single-particle perspective, the new term can be viewed as arising from the different strengths of the curvature and gradient drifts. 
To see this, consider the vacuum-form of the curvature and gradient drifts, given for instance in \textcite{chen2012introduction} by:
\beq
	\vv_{R i} + \vv_{\nabla B i} = \frac{m_i}{Z_i e} \frac{\mathbf{R_c} \times \Bv}{R_c^2 B^2} \lp v_{\parallel i}^2 + \frac{1}{2} v_{\perp i}^2 \rp,
\eeq 
where $\mathbf{R}_c$ is the curvature radius of the magnetic field.
Here we see that for two particles with same total energy, the particle with $v_{\parallel i} > v_{\perp i}$ will have a larger drift.
This drift asymmetry means that even in the absence of a scalar pressure gradient, the pressure anisotropy can still influence the drift speed, and thus create a friction force between species with different drifts.

It is the $F \times B$ drift from this friction force that produces radial magnetized transport. 
Thus, as long as there is a friction force, there will be transport, and the steady state requires the elimination of this diamagnetic friction.
Thus, if one species has an imposed, non-negligible anisotropy and the other does not, the steady state will require a density gradient to form in each species, to counteract the diamagnetic friction due to the anisotropy.
This fundamental need for cancellation within the diamagnetic drift terms will underly the effects that we later derive.

The radial velocity can be calculated by plugging $v_{s\wedge}^{(0)}$ into the radial component of Eq. (\ref{eq:v1s}):
\begin{align}
v_{sr}^{(1)} &= \frac{1}{\Omega_s \tau_{sb}} \lp  v_{s\wedge}^{(0)} \rp + \frac{1}{\Omega_s} \lp \tot{v_{s\wedge}^{(0)}}{t} - b_z^2 k  v_{s\wedge}^{(0)} \rp .   \label{eq:v1s2} 
\end{align}
Here, the first term on the RHS is the friction-driven magnetized transport, i.e. the random walk diffusion of ions across the magnetic field.
The second term on the RHS arises from inertial effects due to compression and rotation, which will have a smaller effect on the dynamics in the regimes of interest.

In order to proceed from this point, we will need to calculate both the total time derivative of the azimuthal velocity and the pressure anisotropy.
Because it will turn out to be less important, we relegate the calculation of the time derivative to Appendix \ref{sec:totalTimeDeriv}, and focus on the pressure anisotropy.

\section{Anisotropy generation} \label{sec:anisotropy}

Consider now the generation of anisotropy in the compressing plasma.
During compression, the plasma will obey the continuity equation and the flux-freezing equation, giving \cite{kulsrud2005plasma}
\begin{align}
n &= n_0 \lp\frac{r}{r_0}\rp^{-2} \label{eq:nScale}\\
B_z &= B_{z0} \lp\frac{r}{r_0}\rp^{-2} \label{eq:BzScale}\\
\Bth &= B_{\theta 0} \lp\frac{r}{r_0}\rp^{-1} \label{eq:BthScale}.
\end{align}
Because the axial magnetic field is associated with a circular area which scales as $r^{2}$, while the azimuthal field is associated with a rectangular area that compresses as $r$, flux freezing ensures that the axial field increases faster than the azimuthal field during compression.
However, the compression leaves the magnetic rotational transform parameter $\kappa$ constant.

We will assume that the pressure evolves according to the double adiabatic MHD equations, given by \cite{kulsrud1980mhd}:
\begin{align}
\tot{}{t} \lp \frac{\pp}{n B} \rp &= 0 \label{eq:DApp}\\
\tot{}{t} \lp \frac{\pl B^2}{n^3} \rp &= 0 \label{eq:DApl}.
\end{align}
Here, Eq.~(\ref{eq:DApp}) is related to conservation of $\mu = m v_\perp^2 / 2|B|$, while Eq.~(\ref{eq:DApl}) is related to conservation of $J = \oint mv_\parallel ds$.

During this compression, let the isotropization proceed according to a species-dependent collision operator:
\begin{align}
\tot{\pp}{t}|_\text{coll} &= \frac{1}{2 \tau_i} \lp \pl - \pp \rp\\
\tot{\pl}{t}|_\text{coll} &= \frac{1}{\tau_i} \lp \pp - \pl \rp,
\end{align}
Here, the single subscript $i$ on $\tau$ denotes the isotropization time of species $i$, in contrast to the double subscript $ij$ from earlier, which denotes the momentum transfer collision time of species $i$ on species $j$.

Combining the adiabatic and relaxation models yields:
\begin{align}
\tot{\pp}{t} &= \frac{\pp}{n B} \tot{}{t} \lp n B \rp + \frac{1}{2 \tau_i} \lp \pl - \pp \rp \label{eq:ppScrew}\\
\tot{\pl}{t} &= \frac{\pl}{n^3 B^{-2}} \tot{}{t} \lp n^3 B^{-2} \rp + \frac{1}{\tau_i} \lp \pp - \pl \rp \label{eq:plScrew}.
\end{align}

We can solve these equations by orders in $\tau_i d/dt \ll 1$.
To $0$th order, Eqs. (\ref{eq:ppScrew}-\ref{eq:plScrew}) reduce to:
\beq
\pl^{(0)} = \pp^{(0)} \equiv p.
\eeq

Then, by taking a linear combination of Eqs. (\ref{eq:ppScrew}-\ref{eq:plScrew}), we solve for the evolution of the 0th-order pressure:
\begin{align}
\tot{p}{t} &\equiv \frac{2}{3} \tot{\pp^{(0)}}{t} + \frac{1}{3} \tot{\pl^{(0)}}{t} \\
&= \frac{2}{3}\frac{\pp^{(0)}}{n B} \tot{}{t} \lp n B \rp + \frac{1}{3} \frac{\pl^{(0)}}{n^3 B^{-2}} \tot{}{t} \lp n^3 B^{-2} \rp \\
&= p\frac{5}{3} \frac{1}{n}\tot{n}{t} \\
&= \frac{10}{3} k p \label{eq:pScale}.
\end{align}
To 0th order, the plasma adiabatically compresses in 3 dimensions, i.e. $p \sim n^{\gamma}$ with $\gamma = 5/3$.

The anisotropy is then given from the first order of Eq. (\ref{eq:ppScrew}), which can be put into the form
\beq
\frac{1}{2 \tau_i} \lp \pp^{(1)} - \pl^{(1)} \rp =  \frac{\pp^{(0)}}{n B} \tot{}{t} \lp n B \rp - \tot{\pp^{(0)}}{t},
\eeq
Expanding $p^{(0)} = n T^{(0)}$ and rearranging yields:
\beq
\eps_i \equiv \frac{\pp^{(1)} - \pl^{(1)}}{p} = 2 k \tau_i \lp \frac{1}{B}\tot{B}{t} - \frac{1}{T^{(0)}}\tot{T^{(0)}}{t} \rp.
\eeq

Taking $B = \sqrt{B_z^2 + \Bth^2}$, and plugging in the scaling relations Eqs.~(\ref{eq:nScale}-\ref{eq:BthScale}) and Eq.~(\ref{eq:pScale}), we thus find
\beq
\eps_i = 2 k \tau_i \lp b_z^2  - \frac{1}{3} \rp . \label{eq:anisotropy}
\eeq
Thus the sign of the anisotropy $\eps_i$ depends on the magnetic geometry.
When $B_z \gg \Bth$, $\eps_i \approx \frac{4}{3} k \tau_i$, while when $B_z \ll \Bth$, $\eps_i \approx -\frac{2}{3} k \tau_i$.

We can understand this result by considering the adiabatic invariants $\mu$ and $J$. 
When $B_z \gg B_\theta$, then we have the relatively strong scaling $|B| \sim r^2$, and $\mu$ conservation rapidly heats the perpendicular degrees of freedom during compression.
Simultaneously, the path length parallel to the (axial) magnetic field remains the same, so that the parallel degree of freedom is not heated.
Conversely, when $B_z \ll B_\theta$, we have the relatively weak scaling $|B| \sim r$, so the perpendicular degrees of freedom are heated less than in the axial-field case.
Furthermore, the path length parallel to the (azimuthal) magnetic field is now a circle, and thus scales as $r$.
Thus the parallel degree of freedom is strongly heated, while the perpendicular degrees of freedom are weakly heated.


The isotropization time $\tau_i$ will in general be species-dependent.
To see this, consider a trace impurity $s$ and a bulk ion $b$, both colliding with the bulk $b$, and with $m_s > m_b$, 
Then, simply from the perpendicular diffusion scalings:
\begin{align}
\tau_{s} &\sim Z_b^{-2} Z_s^{-2} m_b^{-1/2} m_s\\
\tau_{b} &\sim Z_b^{-2} Z_b^{-2} m_b^{-1/2} m_b,
\end{align}
so 
\beq
\frac{\tau_{s}}{\tau_{b}} \sim \lp \frac{Z_b}{Z_s} \rp^2 \lp \frac{m_s}{m_b} \rp. \label{eq:nu}
\eeq
Thus, for $Z_s > Z_b$, we will generally have $\eps_s \leq \eps_b$; i.e. less anisotropy in the impurity than the bulk.

\section{Quasi-stationary state} \label{sec:quasistationary}

We now have all the equations we need to calculate the transport that results from the naturally-produced pressure anisotropy.
Plugging our expressions for anisotropy (Eq.~(\ref{eq:anisotropy})) and inertia (Appendix \ref{sec:totalTimeDeriv}) into our radial velocity (Eq.~(\ref{eq:v1s2})), we find
\begin{align}
\vv_{sr}^{(1)} &= - \frac{T_\perp}{ m_s \Omega_s^2 \tau_{sb}} \biggl\{ \left[ 1 + k \tau_{sb} \lp \frac{4}{3} - 2 b_z^2 \rp\right ] \notag\\
& \times \biggl[ 
\lp \frac{1}{n_s} \pa{n_s}{r} - \frac{Z_s}{Z_b} \frac{1}{n_b} \pa{n_b}{r} \rp  + \lp \eps_s - \frac{Z_s}{Z_b} \eps_b \rp  \frac{b_\theta^2}{r}
\biggr] \biggr\} \label{eq:vrad}\\
\eps_i &= 2 k \tau_i \lp b_z^2  - \frac{1}{3} \rp . \label{eq:anisotropyRep} 
\end{align}

\begin{figure}[t]
	\center
	\includegraphics[width=\linewidth]{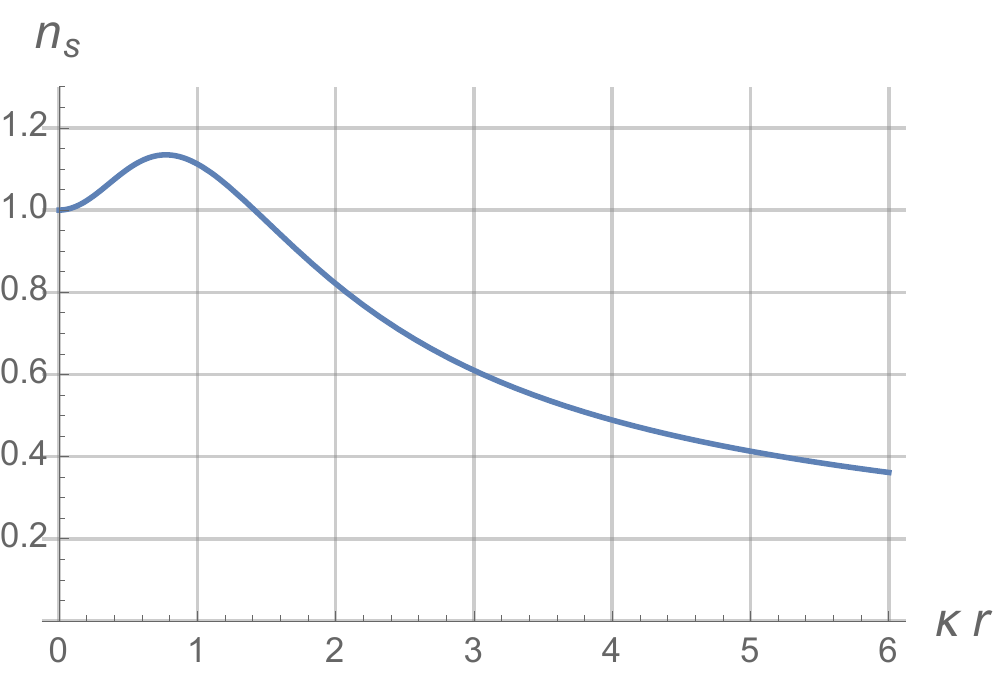}
	\caption{The quasi-stationary state of a compressing screw pinch, i.e. Eq. (\ref{eq:quasiStationary}), as a function of $\kappa r$. Here $k \tau_b = 0.2$, $Z_s / Z_b = 5$, and $m_s / m_b = 10$.
		Impurities will tend to peak at $\kappa r = 0.6$.}
	\label{fig:quasiStationary}
\end{figure}

Before solving for the dynamics of the system, which will involve transforming our fluid equations to a compressing frame, it is helpful to consider the \emph{quasi-stationary state} of the system; i.e. the state in which the radial transport velocity (Eq. \ref{eq:vrad}) vanishes.
Although this is not a true stationary state of the full system of equations, since the anisotropies will tend to change over the course of the compression, it does represent the \emph{state towards which the diffusion equation is evolving on the characteristic diffusion timescale of the system}.
Thus the quasi-stationary state qualitatively reflects the tendency of the impurity to accumulate at the core or edge, and the strength with which it will accumulate.

To get a global solution for the impurity distribution, we must choose a specific form for the magnetic field.
For simplicity, we will consider a screw pinch with a constant rotational transform parameter $\kappa$, defined by
\beq
\kappa \equiv \pa{\theta}{z} = \frac{\Bth(r)}{r B_z(r)}.
\eeq
This setup, in the $P \rightarrow 0$ limit, is the basis for the well-known Gold-Hoyle flux tube \cite{gold1960origin}.
In such a field, the $z$ component of $\hat{b}$ is given by
\begin{align}
b_z &= \frac{1}{\sqrt{1 + \kappa^2 r^2}} \label{eq:bz}\\
b_\theta &= \frac{\kappa r}{\sqrt{1 + \kappa^2 r^2}}. \label{eq:bth}
\end{align}
Note that the scaling Eqs.~(\ref{eq:BzScale}-\ref{eq:BthScale}) imply that $\kappa$ is conserved during the compression.
For further simplicity, we will also assume that the initial magnetic strength is spatially homogeneous.

\begin{figure}[t]
	\center
	\includegraphics[width=\linewidth]{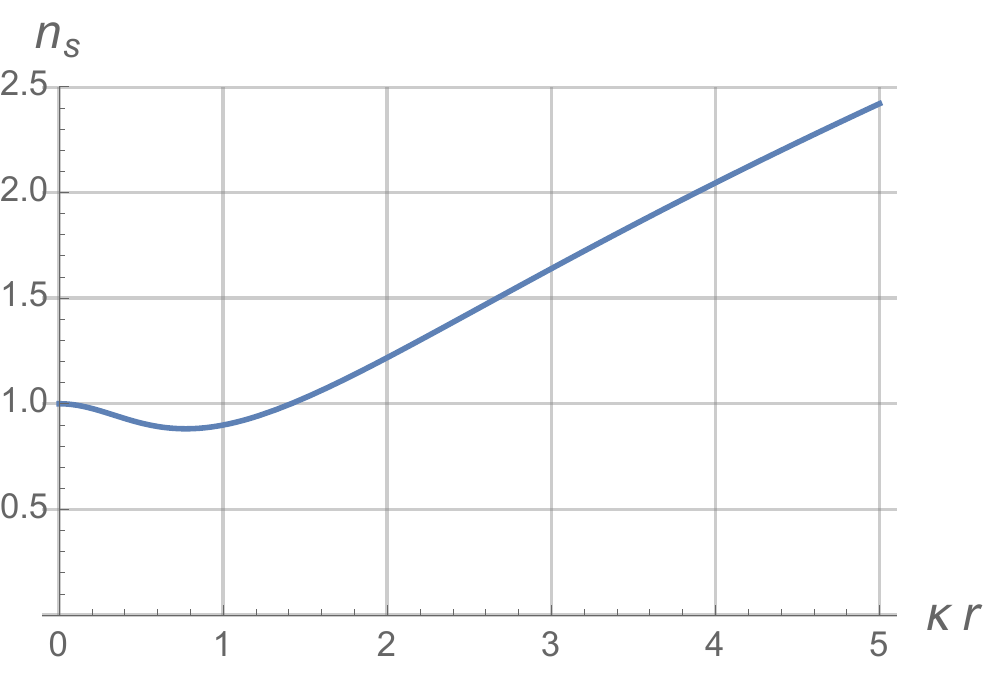}
	\caption{The quasi-stationary state of an expanding screw pinch, i.e. Eq. (\ref{eq:quasiStationary}), as a function of $\kappa r$. Here $k \tau_b = -0.2$, $Z_s / Z_b = 5$, and $m_s / m_b = 10$.}
	\label{fig:quasiStationaryExpand}
\end{figure}

Consider for now the case of flat bulk plasma density.
By setting the left hand side of Eq.~(\ref{eq:vrad}) to 0 and plugging in our magnetic field shape [Eqs.~(\ref{eq:bz}-\ref{eq:bth})], we find
\begin{equation}
n_s(r) \propto \lp 1 + (\kappa r)^2 \rp^{ \frac{1}{2} \lp \frac{Z_s}{Z_b} \eps_b - \eps_s \rp}  \label{eq:quasiStationaryAnis}.
\end{equation}
Thus, even when the scalar density and temperature are flat, the gradient in the pressure anisotropy leads to gradients in the impurity distribution--a qualitatively new classical transport effect.
This anisotropy-dependent transport effect grows stronger as $\kappa r$ grows larger.
Furthermore, since usually $\eps_s \ll \eps_b$, positive anisotropy ($\pps{b} > \pls{b}$) will result in impurity expulsion, while negative anisotropy will result in impurity peaking in the core.

Although the impurity transport direction depends only on the sign of the pressure anisotropy, the anisotropy itself depends on zeroth-order radial velocity (through $k$), and the field pitch magnitude $\kappa$. 
To see this, we simply plug in our expression for anisotropy [Eq.~(\ref{eq:anisotropyRep})] and our anisotropization rate scaling [Eq.~(\ref{eq:nu})] into Eq.~(\ref{eq:quasiStationaryAnis}):
\begin{equation}
n_s(r) \propto \lp 1 + (\kappa r)^2 \rp^{ \frac{Z_s}{Z_b} k \tau_{b} \lp \frac{1}{1 + (\kappa r)^2} - \frac{1}{3} \rp  \lp 1 - \lp \frac{Z_b}{Z_{s}} \rp^3 \frac{m_s}{m_b} \rp } \label{eq:quasiStationary}.
\end{equation}

According to Eq. (\ref{eq:quasiStationary}), in a compressing pinch ($k > 0$), $n_s(r)$ initially increases with radius, achieving a maximum of around $1.147^{k \tau_{b} Z_s / Z_b}$ at $\kappa r = 0.60$, before decaying to 0 in the large-$\kappa r$ limit (Fig.~\ref{fig:quasiStationary}).
Thus, if $\kappa a < 0.6$, impurities will tend to be flushed outwards, while if $\kappa a > 0.6$, they will peak around an interior maximum.
Conversely, in an expanding plasma, anisotropy is driven in the opposite direction, and so too is the transport (Fig.~\ref{fig:quasiStationaryExpand}).

\section{Dynamics in the compressing frame} \label{sec:dynamics}

Our bulk ion population distribution is governed by the equation
\begin{align}
\pa{n_b}{t} &= - \frac{1}{r} \pa{}{r} \lp r n_b v_{br} \rp = \frac{1}{r} \pa{}{r} \lp k r^2 n_b \rp.
\end{align}
If we define our variables in terms of new, scaled variables:
\begin{align}
n_b (r,t) &= n_{b0} \tn (\tr , \tt) e^{2 \tt}\\
\tr &= \frac{r}{a} e^{kt} \label{eq:tr}\\
\tt &= k t,
\end{align}
then our differential equation becomes simply:
\beq
\pa{\tn_b}{\tt} = 0.
\eeq
Thus our background distribution in the new coordinates is simply frozen in place.

We can make the same transformation with our impurity distribution.
We start with
\begin{align}
\pa{n_s}{t} &= - \frac{1}{r} \pa{}{r} \lp r n_s v_{sr} \rp = \frac{1}{r} \pa{}{r} \lp k r^2 n_s - r n_s v_{sr}^{(1)}\rp.
\end{align}

After a change of variables, this becomes
\begin{align}
e^{2 \tt} \pa{\tn}{\tt} &= - \frac{e^{\tt}}{a \tr } \lp \pa{\tr}{r} \pa{}{\tr} \rp \lp a e^{\tt} \tr \tn_s v_{sr}^{(1)} \rp\\
\pa{\tn}{\tt} &= -  \frac{1}{\tr} \pa{}{\tr} \lp \frac{e^{\tt}}{a} \tr \tn_s v_{sr}^{(1)} \rp.
\end{align}

Now, our scaling equations (Eqs.~(\ref{eq:nScale}-\ref{eq:BthScale}) and (\ref{eq:pScale})) combined with our scaled radial coordinate in Eq. (\ref{eq:tr}) imply:
\begin{align}
B_z(\tr) &= B_{z0}(\tr) e^{2 \tt} \label{eq:nScaleC}\\
\Bth(\tr) &= B_{\theta 0}(\tr) e^{\tt}\\
T^{(0)} &= T_0 e^{\frac{4}{3}\tt}\\
\tau_{sb} (\tr) &= \tau_{sb0} (\tr) \label{eq:tauScaleC}.
\end{align}

These combine with the specific form of our field to allow us to calculate the magnetic field unit vectors.
Define $K \equiv \kappa a$.
Then:
\begin{align}
b_z(\tr) &= \frac{B_z(\tr)}{B} = \frac{1}{\sqrt{1 + K^2 \tr^2 e^{-2\tt}}}\\
\bth(\tr) &= \frac{\Bth(\tr)}{B} = \frac{K \tr}{\sqrt{e^{2\tt} + K^2 \tr^2}} \label{eq:bthScaleC}
\end{align}
Note that as $\tt \rightarrow \infty$, $b_z \rightarrow 1$ and $\bth \rightarrow 0$, i.e. the magnetic field becomes straighter.

The above scalings allow us to translate most of the terms in our first-order radial velocity.
The final piece is the inertial term $d v_{s \wedge}^{(0)} / dt$.
Because this term is dominated by the contribution from $E_r$, which depends on the background diamagnetic drift (see Eq.~(\ref{eq:Er})), to 0th order this will have the form;
\beq
\tot{v_{s \wedge}^{(0)}}{t} \approx v_{s \wedge}^{(0)} k \lp \frac{4}{3} - b_z^2 \rp,
\eeq
where we have made use of the above scalings in the explicit form of $v_{s \wedge}^{(0)}$ (Eq.~(\ref{eq:v0})), and assumed that changes in the background pressure due to transport were slow.

When we plug in our radial pinch velocity from Eq.~(\ref{eq:vrad}), and also plug in the above scalings, we find
\begin{align}
\pa{\tn_s}{\tt} &= \frac{1}{\tr} \pa{}{\tr} \biggl\{ \tr \tn_s D_0(\tr) e^{\frac{4}{3} \tt } \lp \frac{1 + K^2 \tr^2}{e^{2 \tt} + K^2 \tr^2 } \rp \notag\\
& \qquad    \times \left[ 1 + k \tau_{sb0}(\tr) \lp \frac{4}{3} - \frac{2}{1+K^2 \tr^2 e^{-2 \tt}}\rp\right ] \notag\\
&\qquad \times \biggl[ 
\lp \frac{1}{\tn_s} \pa{\tn_s}{\tr} - \frac{Z_s}{Z_b} \frac{1}{\tn_b} \pa{\tn_b}{\tr} \rp \notag\\
& \qquad  \qquad + \lp \eps_s - \frac{Z_s}{Z_b} \eps_b \rp \lp \frac{K^2 \tr}{e^{2 \tt} + K^2 \tr^2} \rp
\biggr] \biggr\},
\end{align}
where
\begin{align}
D_0 (\tr) &= \frac{\rho_{s0}^2 (\tr)}{a^2} \frac{1}{k \tau_{sb0}(\tr)}\\
K &= \kappa a.
\end{align}

The first term in the brackets on the second line represents diffusion, and the second represents drive due to anisotropy.
Here, $D_0 (\tr)$ is the initial classical diffusion coefficient as a function of radius, and $\tau_{sb0} (\tr)$ and $\rho_{s0} (\tr)$ are the initial collision frequency and Larmor radius as a function of radius.

Note that the equations at this point are yet fully specified as a function of the densities $\tn_b$ and $\tn_s$.
Although we have specified the ratio $\kappa = \Bth / r B_z$, we have not yet specified the initial magnetic field profile $|B(r)|$.
For our initial study, we take $|B(r)|$ to be constant; this choice ensures that that the diffusion coefficient only varies by a factor of $\mathcal{O}(1)$ across the plasma.
From our scaling Eqs.~(\ref{eq:nScaleC}-\ref{eq:bthScaleC}), we thus find
\beq
D_0 (\tr) = D_{0a} \frac{\tn_{b0}(\tr)}{\tn_{b0a}},
\eeq
where 
\beq
D_{0a} = \frac{\rho_s^2}{a^2} \frac{1}{k \tau_{sb}} |_{\tr = 1,\tt=0}
\eeq
is the initial normalized diffusion coefficient at $\tr = 1$, i.e. at $r = a$.

\begin{figure*}[t]
	\center
	\includegraphics[width=\linewidth]{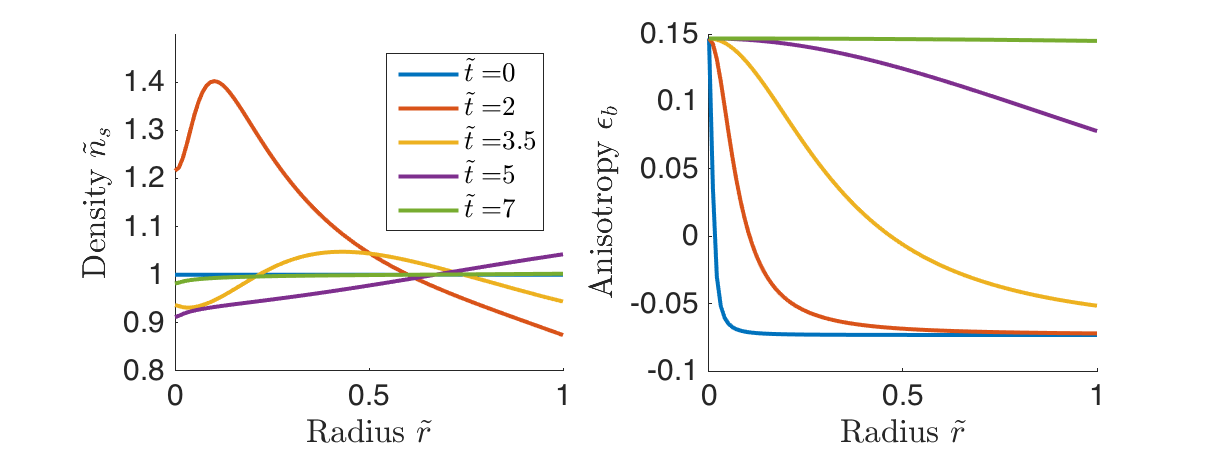}
	\caption{Simulation of anisotropy pinch effect in a flat background plasma.
		Here, $K = 100$, indicating that the plasma initially has large $\Bth$ relative to $B_z$.
		Initially, the plasma compression results in $\pl < \pp$, driving impurities inward, as can be seen at $\tt = 2$, where a peak 40\% higher than the average density forms at $\tr \approx 0.15$.
		Around $\tt = 3.5$, starting at small $\tr$, $B_z$ grows larger than $\Bth$ in a significant part of the plasma interior, and the anisotropy reverses at low-$\tr$.
		By $\tt = 5$ the anisotropy has reversed sign throughout the plasma, driving impurities outward.
		Finally, around $\tt = 7$, the ever-straightening magnetic field makes effects due to anisotropy negligible, and the impurity distribution flattens out.}
	\label{fig:DynamicOsc}
\end{figure*}


Finally, we translate our anisotropy result Eq. (\ref{eq:anisotropy}) to our normalized coordinates, taking account of our scaling relations--this is fairly straightforward.
The final result, putting everything together, is
\begin{align}
\pa{\tn_s}{\tt} &= \frac{1}{\tr} \pa{}{\tr} \biggl\{ \tr \tn_s D_{0a}e^{\frac{4}{3} \tt } \lp \frac{1 + K^2 \tr^2}{e^{2 \tt} + K^2 \tr^2 } \rp   \frac{\tn_{b0}(\tr)}{\tn_{b0a}} \notag \\
&  \qquad \times \left[ 1 + k \tau_{sb0a} \frac{\tn_{b0}(\tr)}{\tn_{b0a}} \lp \frac{4}{3} - \frac{2}{1+K^2 \tr^2 e^{-2 \tt}}\rp\right ] \notag\\
& \qquad \times \biggl[ 
\lp \frac{1}{\tn_s} \pa{\tn_s}{\tr} - \frac{Z_s}{Z_b} \frac{1}{\tn_b} \pa{\tn_b}{\tr} \rp \notag \\
& \qquad \qquad + 
\lp \eps_s - \frac{Z_s}{Z_b} \eps_b \rp \lp \frac{K^2 \tr}{e^{2 \tt} + K^2 \tr^2} \rp
\biggr] \biggr\} \label{eq:Dfinal} \\
\eps_s &= 2 k \tau_{s0a} \frac{\tn_{b0a}}{\tn_{b0}(\tr)} \lp \frac{1}{1 + K^2 \tr^2 e^{-2\tt}} - \frac{1}{3} \rp\label{eq:anisFinal}\\
\eps_b &= \frac{Z_s^2}{Z_b^2} \frac{m_b}{m_s} \eps_s, \label{eq:anisfinal}
\end{align}
where we have defined $\tn_{b0a} \equiv \tn_{b0}(\tr=1)$.
Eqs. (\ref{eq:Dfinal}-\ref{eq:anisfinal}) are the full evolution equations for magnetized transport in the contracting system, including the effects of anisotropy produced by the compression.
In addition to the initial profiles $\tn_b$ and $\tn_s$, the system evolution is determined by 3 dimensionless parameters, evaluated initially at $r = a$.
These are the anisotropy generation $k \tau_{sb0a}$, the diffusion $D_{0a} \equiv \frac{\rho_s^2}{a^2} \frac{1}{k \tau_{sb}}$, and the field line curvature $K \equiv \kappa r$. 

Several other dimensionless parameters must also be checked to ensure that the model is valid, which we derive in Appendix \ref{sec:smallparams}.
These small parameters correspond to the \textbf{C}ollisionality, the \textbf{M}ach number of compression, the \textbf{L}armor radius, and the \textbf{A}nisotropy, all being small compared to relevant scales:
\begin{align}
C_i &\equiv \frac{1}{\Omega_i \tau_{ib}}\\
M &\equiv \frac{|k L|}{v_{th b}}\\
L &\equiv \frac{\rho_s}{L}\\
A_i &\equiv k \tau_{i},
\end{align}
The only parameter which grows over the course of a compression is $L$, which attains a maximum value
\beq
L_{\max} = L \frac{2^{1/3}}{\sqrt{3}} \lp 1+\frac{1}{K^2}\rp^{1/2} K^{1/3} \approx 0.72 K^{1/3}. 
\eeq


\subsection{Features of the Dynamics}

Several features are apparent from the governing equations (\ref{eq:Dfinal}-\ref{eq:anisfinal}).
First, over the course of the shot, the anisotropy drive in the diffusion equation will become less significant, due to the $e^{2\tt}$ term in the denominator of the second term in brackets in Eq. (\ref{eq:Dfinal}).
Thus anisotropy effects occur on an intermediate timescale, before the magnetic field reaches a point where $B_z \gg \Bth$.

A second interesting feature of the equations is the sign of the anisotropy in Eq. (\ref{eq:anisFinal}).
At early times during a compression, the high value of $K$ ensures that $\eps_b < 0$.
However, as the compression continues, $B_z$ becomes larger, until at $B_z = \Bth / \sqrt{2}$ the sign of the anisotropy flips.
At this point, the anisotropy terms will actually try to force the impurities outward, until $B_z$ becomes so large that the anisotropy terms become negligible.

\subsection{Numerical Simulations}

The parabolic Eqs. (\ref{eq:Dfinal}-\ref{eq:anisfinal}) can be solved using Matlab's \texttt{pdepe} function, with reflecting boundary conditions.
Consider a distribution with $\tn_s$ and $\tn_b$ both constant in $\tr$, 
recalling that $\tn_b$ does not change throughout the compression.
Take $Z_s = 5$, $m_s = 10$ for the impurity, and $Z_b = 1$, $m_b=1$ for the background.
Further consider the compression of a fairly low-density, high-field implosion plasma, with $B = 5$ T, $T = 200$ eV, $n = 3 \times 10^{16}$ cm$^{-3}$, $k^{-1} = 2$ $\mu$s, $a=1$ cm, and $K = 100$.
For these parameters, our small parameter orderings are satisfied, with $C_s = 0.06$, $M = 0.036$, $L_{\max} = 0.06$, and $A_b = 0.11$.

The results of this simulation, for $0 < \tt < 7$ and $0 < \tr < 1$, are shown in Figure \ref{fig:DynamicOsc}.
Initially, the anisotropy is negative, i.e. $\pl > \pp$, which drives the impurities inwards.
This impurity accumulation is significant, resulting in an interior peak $40\%$ higher than the average density.
However, around $\tt = 3.5$, enough compression occurs that $B_z > \Bth$ in most of the plasma; thus the sign of the anisotropy reverses, which starts to drive the impurities back out.
This reversal results by $\tt = 5$ in an impurity distribution which is $\sim 15\%$ less concentrated in the core than the edge.
Finally, by $\tt = 7$, $B_z$ becomes so large that effects due to anisotropy are negligible, and the distribution flattens back out due to normal classical diffusion.

These results are not unique to this parameter set. 
For instance, extremely similar dimensionless dynamics are obtained for $B = 3$ T, $T = 20$ eV, $n = 5 \times 10^{15}$ cm$^{-3}$, $k^{-1} = 500$ ns, $a=1$ cm, and $K = 100$, for 3 times ionized Argon ($m_s = 40$) interacting with singly-ionized helium ($m_s = 4$), in the range of several experimental gas-puff experiments\cite{atoyan2016helical,mikitchuk2014mitigation,mikitchuk2018effects}.
However, the small-parameter orderings are more marginally satisfied for this case.

\section{Summary and Discussion}

In this paper, we derived how pressure anisotropy in a cylindrical plasma with a non-negligible azimuthal magnetic field can lead to new impurity transport dynamics.
These dynamics, in the low-pressure limit, can be interpreted as arising from the difference in strength between the ($\pl$-dependent) curvature drift and the ($\pp$-dependent) $\nabla B$ drift.
This difference in drift strengths results in a relative drift velocity (and hence a friction force) between species with different ratios of $\pp$ to $\pl$, which in turn results in radial $F \times B$ drifts that constitute the transport motion.
In particular, a plasma with $\pp > \pl$ tends to flush highly charged impurities to the plasma periphery, while a plasma with $\pp < \pl$ tends to draw in highly charged impurities to the plasma center.

We then showed how pressure anisotropy can be naturally generated in a compressing cylindrically-symmetric plasma, as different adiabatic invariants are conserved parallel and perpendicular to the magnetic field.
The sign of the generated anisotropy (i.e. whether $p_\parallel$ or $p_\perp$ is larger) depends on the relative strength of the axial and azimuthal fields, and on the direction of compression or expansion.
In a compressing plasma with $B_z \ll B_\theta$, the pressure will satisfy $p_\parallel > p_\perp$, which will pull highly-charged impurities into the plasma.
Conversely, when $B_z \gg B_\theta$, $p_\parallel < p_\perp$, which will flush the highly-charged impurities outwards.
The anisotropy, and thus the direction of impurity transport, reverses for an expanding plasma.

Because the axial field increases relative to the azimuthal field over the course of the compression, the impurities will thus tend to be drawn towards the plasma core at short times, before being flushed to the edge at long times.
These effects can be significant, with changes in impurity density on the order of $40\%$ observed for low-density Z-pinch-like parameters.

Our model relied on several simplifying assumptions.
We neglected the anisotropy-dependence of the isotropization $\tau_i$ and frictional collision time $\tau_{ib}$.
Fortunately, any correction to the isotropization time would only show up in our analysis to next order in the anisotropy parameter.
While corrections to the frictional collision rate could have a non-negligible impact on our governing equations, any resulting changes would only show up as an overall (local) multiplicative factor to the diffusion coefficient $D_0(\tr)$, and so would not affect the quasi-stationary state (Eq.~(\ref{eq:quasiStationary})).
Thus, while there could be $O(\eps_b)$ corrections to the diffusion time, there would not be any substantial difference in the core results.

We have also assumed throughout the derivation that the trace impurities are heavier and more highly charged than the bulk plasma ions.
Care must therefore be taken when applying our results to heavy plasmas (such as Argon) with light liners (such as Carbon), or when impurities form a large fraction of the plasma.
While the core results will remain similar, with the more highly-charged species tending to isotropize more quickly, the particulars of the isotropization rate scaling (Eq.~(\ref{eq:nu})) will change.
This change will propagate through to the diffusion equation and quasi-stationary state (Eq.~(\ref{eq:quasiStationary})).

Because the electric and magnetic field evolution were not considered, the analysis was not fully self-consistent; i.e. the form of compression did not represent the actual magneto-hydrodynamic forces on the plasma.
In any real compression, these forces are likely to produce density, temperature, and electromagnetic field gradients which will modify our results, often substantially.
Thus, the current work should be seen not as a realistic calculation of a laboratory compression, but rather as an elucidation of a new process in transport dynamics, which will occur concurrently with other known effects.

A reasonable first step towards a self-consistent calculation would be to make use of known self-similar MHD compression profiles\cite{velikovich1985hydrodynamics}, and to examine the dynamics of impurities within these profiles.
However, such a calculation presents several challenges.
First, it requires the re-inclusion of thermal gradient driven terms such as the Nernst friction, which must be calculated in the presence of finite anisotropy.
Second, it introduces regimes of mixed collisionality, where $\Omega_i \tau_{ib}$ can be greater or less than one depending on the location within the plasma.
Thus, both magnetized and unmagnetized transport will occur in the plasma at different locations, significantly complicating the self-consistent dynamics.

One additional complication is likely to occur as the plasma compresses and the ratio of thermal to magnetic pressure (plasma $\beta$) increases.
At high beta, the anisotropy will eventually be determined not by collisions, but rather by kinetic instabilities such as the mirror and firehose\cite{parker1958dynamical}, which will cause the anisotropy to saturate around $\eps_b \sim 1/\beta$.

Nevertheless, the large effects of anisotropy on the transport should be visible during low-density laboratory plasma implosions, especially as they are pushed towards larger magnetic fields.
The greatest difficulty in experimental analysis is likely to be disentangling these transport effects from those driven by temperature and density gradients, since this requires having accurate radial profiles for the density and temperature.
Thus, understanding the interesting transport consequences of the pressure anisotropy, which can flush out or draw in impurities depending on the field structure, is important in understanding experimental results.

\section*{Acknowledgments}
The authors thank Elijah Kolmes, Mikhail Mlodik, and Seth Davidovits for helpful discussions.
This work was supported by NNSA 83228-10966 [Prime No. DOE (NNSA) DE-NA0003764], and by NSF PHY-1506122. 
One author (IEO) also acknowledges the support of the DOE Computational Science Graduate Fellowship (DOE grant number DE-FG02-97ER25308).


\appendix
\section{Total time derivative of azimuthal velocity} \label{sec:totalTimeDeriv}

Here, we calculate the total derivative of the azimuthal velocity as required by Eq. (\ref{eq:v1s2}).
From Eq. (\ref{eq:v0}), we have:
\beq
\vv_{s\wedge}^{(0)} = \frac{\lp \nabla \cdot \Pv_b \rp \cdot \hat{r}}{n_b Z_b e |B|} - \frac{\lp \nabla \cdot \Pv_s \rp \cdot \hat{r}}{n_s Z_s e |B|}. 
\eeq
During a short period of contraction (compared to the diffusion timescale), we have $r = r_0 e^{-kt}$, $P^{(0)} \ P^{(0)}_0 e^{10 kt/3}$, $n = n_0 e^{2kt}$, and $|B| = \sqrt{B_{z0}^2 e^{4kt} + B_{\theta 0}^2 e^{2kt}}$.
Thus, over a short contraction time,
\begin{align}
\vv_{s\wedge}^{(0)} &= \frac{e^{7kt/3}}{\sqrt{B_{z0}^2 e^{4kt} + B_{\theta 0}^2 e^{2kt}}} \lp \frac{\lp \nabla \cdot \Pv_b \rp \cdot \hat{r}}{n_b Z_b e |B|} - \frac{\lp \nabla \cdot \Pv_s \rp \cdot \hat{r}}{n_s Z_s e |B|} \rp_{t=0}  \\
&= \frac{e^{7kt/3}}{\sqrt{B_{z0}^2 e^{4kt} + B_{\theta 0}^2 e^{2kt}}} \lp \vv_{s\wedge}^{(0)} \rp_{t=0}  . 
\end{align}
Taking the time derivative and evaluating at $t=0$, we find:
\beq
	\frac{dv_{s\wedge}^{(0)}}{dt} = k \lp \frac{4}{3} - 2 b_z^2 \rp \vv_{s\wedge}^{(0)}.
\eeq

\section{Small parameters} \label{sec:smallparams}

Over the course of the derivation, we have collected several parameters which must be small:
\begin{align}
\frac{1}{\Omega_i \tau_{ib}} &\ll 1 \label{eq:constraintMag}\\
\left | \frac{k}{\Omega_s} \frac{k r}{v_{s\wedge}^{(0)}} \right| &\ll k \tau_{sb} \label{eq:constraintMach}\\
\left | b_z^2 \frac{v_{s\wedge}^{(0)}}{\Omega_s r} \right | &\ll k \tau_{sb}	\label{eq:constraintRot}\\
|k \tau_{ib} | &\ll 1 \label{eq:constraintAnis},
\end{align}
where the subscript $i$ indicates that the expression applies to both species $s$ and $b$.
The first represents the requirement of magnetized diffusion, the second and third represent the negligibility of inertial terms, and the last represents the assumption of small anisotropy.
As our system compresses, these parameters will evolve.
Thus we must ensure that we confine our analysis to systems in which our approximations remain valid throughout the compression.

We start by making simpler small parameters that enforce the orderings in Eqs. (\ref{eq:constraintMag}-\ref{eq:constraintAnis}).
To do this, we note that $|\nabla \cdot \Pv_i| \sim n_i T / r$.
Thus, plugging Eq. (\ref{eq:v0}) into Eq. (\ref{eq:constraintMach}) and noting that the background diamagnetic drift term is larger than the impurity diamagnetic drift term, we find
\beq
\left | \frac{k}{\Omega_s} \frac{k r}{v_{s\wedge}^{(0)}} \right| \sim \frac{Z_b}{Z_s}\frac{m_s r^2 k^2}{T}   \sim \frac{m_b r^2 k^2}{T} = \frac{\lp v_r^{(0)}\rp^2}{v_{thb}^2} \equiv M^2,
\eeq
where $M$ is the Mach number of the compressing flow.
Note that in the second-to-last scaling, we assumed that $Z_b / Z_s \sim m_b / m_s$. If $s$ is only weakly ionized, the second to last scaling might not hold, since then $m_b \ll m_s$ but $Z_b \sim Z_s$; thus, when dealing with very different charge to mass ratios, it is best to keep $M_s^2 \ll Z_s / Z_b$.

Similarly, we can express Eq. (\ref{eq:constraintRot}) as 
\beq
\left | b_z^2 \frac{v_{s\wedge}^{(0)}}{\Omega_s r} \right | \sim \frac{\rho_s \rho_b}{r^2} b_z^2 \lesssim \lp \frac{\rho_s}{r} \rp^2.
\eeq
Thus we see that the second inertial constraint is satisfied when we adopt the familiar requirement that the Larmor radius be smaller than the system size.

Our formal small parameters that we adopt correspond to the \textbf{C}ollisionality, the \textbf{M}ach number of compression, the \textbf{L}armor radius, and the \textbf{A}nisotropy, all being small compared to relevant scales:
\begin{align}
C_i &\equiv \frac{1}{\Omega_i \tau_{ib}}\\
M &\equiv \frac{|k L|}{v_{th b}}\\
L &\equiv \frac{\rho_s}{L}\\
A_i &\equiv k \tau_{i},
\end{align}
where $L = a e^{-\tt}$ is the compressing system scale.
Note that, if $M, L \sim A_i \ll 1$, then $M^2, L^2 \ll A_i$, and our constraints Eqs. (\ref{eq:constraintMach}-\ref{eq:constraintRot}) for anisotropic effects to dominate over inertial effects are automatically satisfied.\\


Note also that these four parameters are not fully independent, since
\beq
M L = \sqrt{\frac{m_b}{m_s}} \frac{\tau_{sb}}{\tau_{s}} C_s A_s.
\eeq

Finally, note that anisotropy effects are determined by $ A_b = \frac{Z_s^2}{Z_b^2} \frac{m_b}{m_i} A_s$, and the diffusion coefficient is given by the $\tt = 0$ value of $D_{0a} = (\tau_{s} / \tau_{sb}) L^2 / A_s$.\\

\textbf{Scaling of small parameters:}
As we compress, the small parameters will evolve in a way determined by the scalings in Eqs. (\ref{eq:nScaleC}-\ref{eq:bthScaleC}).
We find:
\begin{align}
C_{i} &\sim \frac{1}{B} \sim e^{-\tt} \lp \frac{1 + K^2 }{e^{2\tt} + K^2 } \rp^{1/2} \\
M &\sim \frac{r}{T^{1/2}} \sim e^{-\frac{5}{3} \tt} \\
L &\sim \frac{T^{1/2}}{r B} \sim e^{\frac{2}{3} \tt} \lp \frac{1 + K^2 }{e^{2\tt} + K^2 } \rp^{1/2}\\
A_i &\sim \text{const}.
\end{align}
As we compress, all our small parameters shrink or stay constant, with the exception of $L$.
Thus we need to choose an initial $L$ small enough such that at no point during the simulation will $L \gtrsim 1$.

A sufficient condition to meet this constraint is to take:
\beq
L < L_{\max} e^{-2\tt_{\max}/3 }. 
\eeq
However, this is in general too stringent, since for a long compression, $S$ should begin to shrink (once $B_z \sim \Bth$).
Indeed, for $K >1$, the maximum value achieved by $S$ for $\tt > 0$ is
\beq
L_{\max} = L \frac{2^{1/3}}{\sqrt{3}} \lp 1+\frac{1}{K^2}\rp^{1/2} K^{1/3} \approx 0.72 K^{1/3}. 
\eeq
Thus, we simply choose
\beq
L =  L_{\max} \max \lp e^{-2\tt_{\max}/3 }, 1.36 K^{-1/3} \rp.
\eeq
This constraint will limit the initial value of our diffusion coefficient significantly.

\section*{References}
%

\clearpage

\end{document}